\documentclass{emulateapj}
\usepackage{apjfonts}
\usepackage{natbib}
\usepackage{epsfig}

\newcommand{\tQ}{t_{Q}}
\newcommand{\dtdl}{d\tQ/d\log(L)}
\newcommand{\dtdL}{\dtdl}
\newcommand{\Lp}{L_{\rm peak}}
\newcommand{\lp}{\Lp}
\newcommand{\nLp}{\dot{n}(\Lp)}

\newcommand{\Lbol}{L_{\rm bol}}
\newcommand{\lbol}{\Lbol}
\newcommand{\Lstar}{L_{\ast}}
\newcommand{\lstar}{\Lstar}
\newcommand{\slope}{\gamma}
\newcommand{\tslope}{\beta}
\newcommand{\slopezero}{\slope_{12}}
\newcommand{\dslope}{\frac{d\slope}{d\log(\Lp)}}

\newcommand{\mbh}{M_{\rm BH}}

\newcommand{\Mdot}{\dot{M}}

\newcommand{\etal}{et al.}
\newcommand{\geos}{\gamma_{\rm EOS}}

\shorttitle{The Faint-End QLF Slope}
\shortauthors{Hopkins \etal}
\slugcomment{Submitted to ApJ, August 5, 2005}
\begin{document}

\title{The Evolution in the Faint-End Slope of the Quasar Luminosity Function}
\author{Philip F. Hopkins\altaffilmark{1}, 
Lars Hernquist\altaffilmark{1}, 
Thomas J. Cox\altaffilmark{1}, 
Brant Robertson\altaffilmark{1}, 
Tiziana Di Matteo\altaffilmark{2}, 
\&\ Volker Springel\altaffilmark{3} 
}
\altaffiltext{1}{Harvard-Smithsonian Center for Astrophysics, 
60 Garden Street, Cambridge, MA 02138, USA}
\altaffiltext{2}{Carnegie Mellon University, 
Department of Physics, 5000 Forbes Ave., Pittsburgh, PA 15213}
\altaffiltext{3}{Max-Planck-Institut f\"{u}r Astrophysik, 
Karl-Schwarzchild-Stra\ss e 1, 85740 Garching bei M\"{u}nchen, Germany}

\begin{abstract}

Based on numerical simulations of galaxy mergers that incorporate
black hole growth, we predict the faint end slope of the quasar
luminosity function (QLF) and its evolution with redshift.  Our
simulations have yielded a new model for quasar lifetimes and light
curves where the lifetime depends on both the instantaneous {\it and}
peak luminosities of the quasar.  This description motivates a new
interpretation of the quasar luminosity function in which the bright
end of the QLF consists of quasars radiating near their peak
luminosities, but the faint end is mostly made up of brighter peak luminosity quasars seen in
less luminous phases of evolution.  The faint-end QLF slope is then
determined by the faint-end slope of the quasar lifetime for quasars
with peak luminosities near the observed break.  We determine this
slope from the quasar lifetime as a function of peak luminosity, based on 
a large set of simulations which encompass a wide variety of host galaxy, 
merger, black hole, and interstellar gas properties.
Brighter peak luminosity (higher black hole mass) systems undergo more
violent evolution, and expel and heat gas more rapidly in the final
stages of quasar evolution, resulting in a flatter faint-end slope (as
these objects fall below the observed break in the QLF more
rapidly). Therefore, as the QLF break luminosity moves to higher
luminosities with increasing redshift, implying a larger typical
quasar peak luminosity, the faint-end QLF slope flattens. From our
determined quasar lifetimes as a function of peak luminosity and this
interpretation of the QLF, we predict the faint-end QLF slope and its
evolution with redshift in good agreement with observations.  Although 
black holes grow in an anti-hierarchical manner (with lower-mass black holes 
formed primarily at lower redshifts), the
observed change in slope and differential or ``luminosity dependent
density evolution'' in the QLF is completely determined by the
non-trivial, luminosity-dependent quasar lifetime and physics of
quasar feedback, and not by changes in the shape of the underlying
peak luminosity or active black hole mass distributions.

\end{abstract}

\keywords{quasars: general --- galaxies: active --- 
galaxies: evolution --- cosmology: theory}

\section{Introduction}
\label{sec:intro}

The shape and evolution of the quasar luminosity function (QLF) is
fundamental to cosmology, constraining theories of galaxy and
supermassive black hole (BH) formation, accretion models, the X-ray,
UV, and infrared backgrounds, the role of galaxy mergers and
interactions, and reionization. Until recently, it has not been
possible to reliably measure the faint-end slope of the QLF even at
low redshifts, but this has begun to change with the advent of large,
uniformly selected quasar samples in surveys such as the SDSS and 2dF
which now increasingly probe the QLF to fainter luminosities. Furthermore, a
growing number of observations at different redshifts, in radio,
optical, and soft and hard X-ray bands, have suggested that the faint end
slope evolves, becoming flatter at higher redshift
\citep[e.g.][]{Page97,Miyaji00,Miyaji01,LaFranca02,Cowie03,
Ueda03,Fiore03,Hunt04,Cirasuolo05,HMS05}.

This evolution, parameterized either as direct evolution in the
faint-end slope or as ``luminosity-dependent density evolution''
(LDDE), has been the subject of much speculation, as it implies that
the density of lower-luminosity quasars peaks at lower redshift. In
traditional models of the quasar lifetime and light curve, this
evolution must be directly related to the distribution of quasar
hosts, implying e.g.\ significant and rapid evolution in the {\em
shape} of the distribution of host galaxy masses, which cannot be accounted
for in either semi-analytical models or numerical simulations and is
not consistent with a wide range of galaxy observations.  Although
models which adopt these simplified prescriptions for quasar evolution
have had success predicting the evolution of the {\em bright} end of
the QLF \citep[e.g.][]{KH00,WL03}, they do not predict the faint-end slope or its evolution, and
as such cannot be reliably extrapolated to low luminosities or to
redshifts where the slope is undetermined. Observations of the high-redshift, faint
end quasar luminosity function slope are highly uncertain, and no large, uniformly 
selected samples yet exist which measure the faint-end slope at both low ($z\lesssim1$) and 
high ($z\gtrsim3$) redshifts, and therefore theoretical models of the 
faint-end slope are especially important. This high-$z$, faint-end
slope is a critical quantity in determining the early formation
history of black holes and, especially, their contribution to
reionization, as well as possible connections between quasars and
e.g.\ the low-luminosity Seyferts seen at $z\sim0$.

Without a more sophisticated model of quasar evolution, models which
have attempted to reconcile observations of evolution in the faint-end
QLF slope and BH populations \citep[e.g.,][]{Merloni04} have been
forced to fit to the entire QLF and BH mass distribution to
essentially arbitrary distributions of lifetimes/duty cycles and
accretion rates as a function of redshift. Still, this
phenomenological modeling has elucidated the anti-hierarchical nature
of BH growth, with smaller-mass BHs formed at lower redshift as an
implication of this evolution in the QLF. But an actual prediction of
the faint-end slope requires a more detailed model for both quasar
lifetimes and the QLF.

Recently, BH growth and feedback have been incorporated into numerical
simulations of galaxy mergers (Springel et al.\ 2005a,b, 
Springel 2005, Di Matteo et al.\ 2005).  In these
simulations, gravitational torques drive inflows of gas into the
nuclei of merging galaxies (e.g. Barnes \& Hernquist 1991, 1996),
triggering starbursts (e.g. Mihos \& Hernquist 1996) and feeding the
growth of central supermassive BHs (Di Matteo et al. 2005).  As the
BHs accrete, some of the radiated energy couples to the surrounding
gas, and the growth eventually stalls when this feedback energy is
sufficient to unbind the reservoir of gas.  These
simulations elucidate the connection between galaxy evolution, the
formation of supermassive BHs, and the self-regulated nature of quasar
activity, and provide quantitative predictions which agree well with
observations of, e.g., the $M_{\rm BH} - \sigma$ relation (Di Matteo
et al. 2005, Robertson et al. 2005), quasar lifetimes (Hopkins et
al. 2005a,b), and the QLF in various wavebands (Hopkins et
al. 2005c,d,e).  These simulations provide a fully self-consistent,
quantitative prediction of the light curve of quasar activity and the
resulting quasar ``lifetime'' -- i.e., the amount of time that a
quasar spends at a given luminosity -- and its dependence on the
properties of the merging galaxies.

The resulting quasar light curves imply a qualitatively different
picture of the QLF than previously considered \citep{H05c}.
Specifically, in the model of Hopkins et al. (2005a-f), quasars evolve
rapidly and their lifetime depends on both their instantaneous {\it
and} peak luminosities. Critically, the quasar lifetime in this model
is longer at lower luminosities; i.e.\ a given quasar spends more time
(and is more likely to be observed) at a luminosity well below its
peak luminosity. This differs from previous models which generally
assume that quasars radiate at a fixed luminosity for some
characteristic lifetime, or adopt simplified exponential light curves.
In our picture, the bright end of the QLF consists of quasars growing
rapidly and radiating near their peak luminosities, while the faint
end consists of quasars either undergoing exponential growth to much
larger masses and higher luminosities, or in sub-Eddington quiescent
states going into or coming out of a period of peak activity. The
``break'' in the QLF corresponds directly to the maximum in the
intrinsic distribution of {\it peak} luminosities, which falls off at both
brighter and fainter luminosities. The faint-end QLF slope is then
determined by the faint-end slope of the luminosity-dependent 
quasar lifetime (i.e.\ the differential time the quasar spends in 
any given luminosity interval), for quasars
with peak luminosity near the observed break. In other words, the
number of quasars in efficient, near-peak growth stages needed to
account for the observed bright end of the QLF is already sufficient
to account for the observed faint end, as any given quasar is much
more likely to be observed at luminosities well below its peak, and
thus the probability of seeing such brighter sources at these lower 
luminosities entirely determines the shape of the faint-end QLF.

This modeling of the quasar lifetime, and the corresponding
interpretation of the QLF, then provides a physical motivation for the
location of the break luminosity and the faint-end slope of the
observed QLF \citep{H05c,H05d,H05e}. Furthermore, \citet{H05e}
demonstrate that this description accurately predicts a large number
of quasar observations, including the QLF in all wavebands and at all
redshifts, the distribution of Eddington ratios as a function of
luminosity and redshift, the BH mass function (both of active and
relic BHs), the distribution of column densities and obscured fraction
as a function of luminosity, the X-ray background spectrum, and the
anti-hierarchical BH growth described above. Given this remarkable
agreement and the simple physical motivation for the faint-end QLF
slope unique to this modeling, it should account for both the value
and evolution of this slope with redshift.

In this paper, we use our picture of quasar lifetimes and the resulting
QLF interpretation from Hopkins et al.\ (2005a-f) to predict the
faint-end slope of the quasar lifetime as a function of quasar peak
luminosity, which in this model can be immediately translated to a
prediction of the faint-end QLF slope as a function of the QLF break
luminosity and, consequently, as a function of redshift. We are able
to accurately predict the observed evolution in the slope and
corresponding LDDE based only on the input physics of our hydrodynamic
simulations, as a unique consequence of our interpretation of the QLF
and without the need to invoke any evolution in the shape of e.g.\ the
quasar host or galaxy mass distribution.  We further develop a simple
analytical model of quasar feedback which gives similar predictions,
and demonstrates the scale-dependent feedback physics which drive the
evolution of the faint-end slope.

Throughout, we define $L=\lbol$, the bolometric luminosity. 
We denote the faint-end slope $\slope$ by 
$\phi=d\Phi/d\log(L)\propto L^{-\slope}$, where $\phi$ is 
the differential {\em bolometric} QLF and we are 
considering luminosities $L\ll\lstar$, 
where $\lstar$ is the ``break'' in the QLF. 
Note that $d\Phi/dL\propto L^{-\slope-1}$, as e.g.\ 
in the standard ``double power law'' form of the QLF. The sign convention
(i.e.\ {\em positive $\slope$}) also follows the general observational 
convention and standard double power law convention, with 
$d\Phi/dL\propto1/\{(L/L_{0})^{\gamma_{1}+1}+(L/L_{0})^{\gamma_{2}+1}\}$.
We adopt a $\Omega_{\rm M}=0.3$, $\Omega_{\Lambda}=0.7$, 
$H_{0}=70\,{\rm km\,s^{-1}\,Mpc^{-1}}$ cosmology.

\section{The Faint-End Slope as a Function of Peak Luminosity}
\label{sec:slope}

If quasars spend a differential time $\dtdl$ per logarithmic 
interval in luminosity, then the observed QLF 
is (at times when the quasar lifetime is short compared to the Hubble time)
\begin{equation}
\phi(L)=\int{\dtdl\,\nLp\,d\log\Lp},
\end{equation}
where $\nLp$ is the rate at which quasars of a given peak luminosity $\Lp$ 
are created or activated (per unit time per unit comoving volume per 
logarithmic interval in $\Lp$). In \citet{H05c,H05e}, we use this 
and the realistic modeling of quasar lifetimes from 
\citet{H05a,H05b,H05e} to determine $\nLp$, and find that it does {\em not} 
have the same shape as the observed QLF, as is predicted 
by simple models in which the quasar turns ``on/off'' as a step function 
or follows a pure exponential light curve. Instead, $\nLp$ traces the 
shape of the observed QLF at the bright end (above $\lstar$), peaks at
$\Lp\sim\lstar$, then falls off below this. Therefore, the 
slope of the faint-end QLF, $\slope$, is dominated by the faint-end 
slope of $\dtdl$ for quasars with $\Lp\sim\lstar$;
i.e.\ the observed 
faint end of the QLF is dominated by brighter {\em peak} luminosity sources, accreting 
efficiently but in early merger stages or in sub-Eddington states transitioning 
into or out of peak quasar activity, so $\slope$ is determined by the integrated probability of 
seeing the population of brighter $\Lp$ sources, dominated by sources with 
$\Lp\sim\lstar$, at lower $L$.

Given this interpretation of the QLF, $\slope$ should be simply calculable  
from the $L\ll\Lp$ slope of $\dtdL$ (since the probability of seeing 
such a source at $L$ is proportional to its lifetime at $L$), for 
sources with $\Lp\approx\lstar$. If the quasar lifetime is a function 
of $L$ and $\Lp$ only (i.e.\ not affected systematically by other host galaxy properties), then 
$\slope$ can be entirely predicted by knowing $\lstar(z)$, which directly 
gives the peak in the $\nLp$ distribution as a function of redshift. 
There may be some curvature introduced because of the non-zero contributions 
to the faint-end of the QLF from sources with $\Lp\neq\lstar$, but as such 
corrections depend on the exact shape of $\nLp$ and are small, as 
$\nLp$ drops off rapidly (as the bright-end QLF does) away from $L=\lstar$, 
we ignore them here.  

\subsection{The Quasar Lifetime from Simulations}
\label{sec:sims}

We first consider the $L\ll\Lp$ behavior of the quasar lifetime 
determined from hydrodynamical simulations. We consider
a suite of several hundred simulations, described in detail in \citet{Robertson05}
and \citet{H05e}.
These were performed with
the new parallel TreeSPH code GADGET-2 (Springel 2005),
which uses an entropy-conserving formulation of
smoothed particle hydrodynamics \citep[SPH;][]{SH02}, and includes a
sub-resolution, multiphase model of the dense interstellar medium
(ISM) to describe star formation \citep{SH03}.  The multiphase gas is
pressurized by feedback from supernovae, allowing us to stably evolve
even pure gas disks (see, e.g. Springel et al.\ 2005b, Robertson et al.\ 2004<
Springel \& Hernquist 2005). BHs are
represented by ``sink'' particles that accrete gas, with an accretion
rate $\Mdot$ estimated using a Bondi-Hoyle-Lyttleton parameterization,
with an upper limit equal to the Eddington rate \citep{SDH05b}. The
bolometric luminosity of the BH is then $L=\epsilon_{r}\Mdot c^{2}$, where
$\epsilon=0.1$ is the radiative efficiency.  We further allow a small
fraction (typically $\approx 5\%$) of $L$ to couple dynamically to the gas as
thermal energy. This fraction is a free parameter, determined in
\citet{DSH05} by fitting the $M_{\rm BH}$-$\sigma$ relation.
We generate two stable, isolated disk galaxies, each with an extended
dark matter halo having a \citet{Hernquist90} profile, an exponential
disk, and a bulge. The galaxies are then set to
collide from a zero energy orbit.

In our suite of simulations, we vary the masses and virial velocities of the 
initial galaxies, halo concentrations, ISM equation of state, 
parameters describing feedback from supernovae and black hole growth, 
presence or absence of bulges in the host galaxies, initial BH 
seed masses, numerical resolution 
(typically $\sim2\times10^{5}$ particles per galaxy, but we consider 
up to 10-100 times as many), disk inclinations and pericenter separation of 
the initial orbits, and initial disk gas fractions. We further scale all galaxy 
properties appropriately to resemble galaxies at redshifts $z=0-6$, 
for a large subset of our simulations as described in \citet{Robertson05}. 
Our simulations produce quasars with $L,\ \Lp$ from $\sim10^{8}-10^{15}\,L_{\sun}$, 
spanning the entire range of observed quasar luminosities at all redshifts. 
\citet{H05e} consider this 
set of simulations in detail, and use it to determine the quasar lifetime $\dtdl$. 
Critically, they find that $\dtdl$ shows {\em no} systematic dependence 
and little scatter with any of the varied parameters, when quantified as a 
function of $L$ and $\Lp$. The predictions in \citet{H05e} did not 
depend sensitively on the (much more uncertain) faint-end slope, and therefore the authors 
did not consider a detailed fit to the faint-end behavior 
of $\dtdl$, but rather parameterized the lifetime as a simple 
exponential, $\dtdl=t_{0}\,\exp(-L/L_{0})$, where $L_{0}\approx0.2\Lp$
and $t_{0}$ depends weakly on $\Lp$, 
which provides an excellent fit to the simulation results for 
$L\gtrsim 10^{-2}-10^{-1}\Lp$.  
\citet{H05b} consider the $L\ll\Lp$ behavior of $\dtdl$ in 
more detail, and find that the lifetime more closely resembles a 
power law at these $L$, with a power-law slope $\alpha$ a function of $\Lp$. 
The combination of these results suggests that the quasar lifetime 
is best parameterized as a Schechter function with slope $\alpha$, 
normalization $t_{0}(\Lp)$, and turnover $L_{0}(\Lp)$, for purposes 
where the faint-end slope is important; i.e.\ (to clarify our conventions) 
\begin{equation}
\frac{{\rm d}t_{Q}}{{\rm d}\log(L)}=t_{0}\,{\Bigl(}\frac{L}{L_{0}}{\Bigr)}^{-\alpha}\,
\exp{{\Bigl(}-\frac{L}{L_{0}}{\Bigr)}}.
\end{equation} 
For a QLF with break $\lstar(z)$, which implies an $\nLp$ distribution peaked
at $\Lp\sim\lstar$, the observed faint-end QLF slope is then $\slope\approx\alpha(\Lp=\lstar)$.

Here, we consider the faint-end ($L\ll\Lp$) slope from these simulations 
in greater detail. For each simulation, we fit a Schechter function to the 
quasar lifetime as a function of luminosity. We find similar 
results to \citet{H05e} for $L_{0}$ and $t_{0}$ (allowing for the 
well-known degeneracy between $t_{0}$ and $\alpha$ in these fits), 
but now quantify $\alpha$ as a function of $\Lp$. 
As discussed in \citet{H05c,H05e}, these faint-end slopes are subject 
to several uncertainties in our modeling
(the reason for ignoring the 
correction at very low luminosities implied by $\alpha$ in Hopkins et al.\ 2005e). 
One is the finite time duration of our simulations, which may flatten 
$\dtdl$ at low $L$, as the BH cannot completely relax 
in the given time. A complementary means to determine $\slope$ is then 
to consider the rate at which $L$ falls off after $L=\Lp$ at time $t=t_{p}=t(\Lp)$. 
We take $L(t)/\Lp$ to be a function of $t-t_{p}$, and find 
that it falls off in approximate power-law fashion for all our simulations, 
$L(t)\propto (t-t_{p})^{-\tslope}$. This implies $\dtdl\propto L^{-1/\tslope}$, 
i.e.\ $\slope=\alpha=1/\tslope$. Although these fits are not affected by the finite 
time duration of the simulations, they do not include the time spent 
at different $L$ before $L=\Lp$, and thus are entirely accurate only at $t\gg t_{p}$
or for a symmetric rise/fall in $L$, and may therefore 
overestimate the steepness of $\slope$, so we consider both fitting methods below.

The uncertainties in this modeling at low luminosities are 
discussed in detail by \citet{H05c,H05e}. \citet{H05b} examine
fits to the power-law behavior of $\dtdl$ at $L\ll\Lp$ considering 
both the entire simulation and only times after the merger (similar 
to our fits for the decay of $L(t)$). Further, \citet{H05c} 
consider the application of an ADAF-type correction for 
radiatively inefficient accretion flows at low accretion rates \citep[following e.g.,][]{NY95}, 
to account for more detailed variation in the radiative efficiency and 
spectrum as a function of accretion rate. We apply this correction
in our fitting as well. Although these previous works 
consider a much smaller subset of the simulations we 
have fit to, they find similar results, and 
we compare their fits and ours below to demonstrate both the general agreement 
and range of uncertainty introduced by these different determinations of $\slope$. 

\begin{figure*}
    \centering
    \plotone{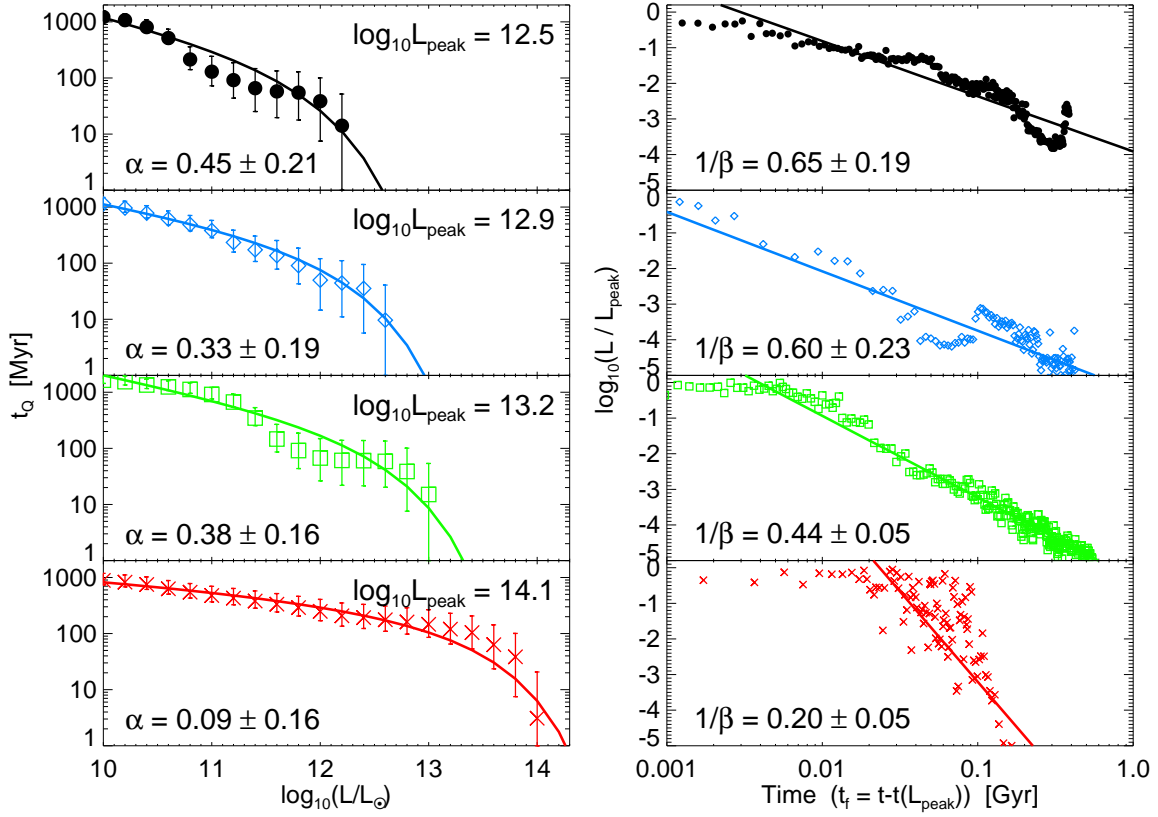}
    \caption{Left: Quasar lifetime (integrated time above a given $L$), for 
    representative mergers with peak luminosities 
    $\Lp/L_{\sun}=3.1\times10^{12}\ {\rm (black\ circles)},\ 7.3\times10^{12}\ {\rm (blue\ diamonds)},\ 
    1.7\times10^{13}\ {\rm (green\ squares)},\ 1.4\times10^{14}\ {\rm (red\ \times's)}$, 
    with the best-fit Shechter functions (lines) of slope $\alpha$. 
    Right: Bolometric luminosity over peak luminosity ($L/\lp$) as a function of 
    time ($t_{f}=t-t(\Lp)$), with the best-fit power law of slope $\tslope$.
    The predicted faint-end LF slopes $\slope=\alpha=1/\tslope$ are also shown.
    \label{fig:sims.ex}}
\end{figure*}

Figure~\ref{fig:sims.ex} shows the results of this fitting for several representative 
simulations. We show the lifetime as a function of $L$
(for clarity, the integrated lifetime above a given 
$L$, $t_{Q}(L'>L)$, is shown),  for simulations 
with $\Lp/L_{\sun}=3.1\times10^{12}\ {\rm (black\ circles)},\ 
7.3\times10^{12}\ {\rm (blue\ diamonds)},\ 
1.7\times10^{13}\ {\rm (green\ squares)},\ 
1.4\times10^{14}\ {\rm (red\ \times's)}$, and the best-fit Schechter 
function (solid lines) to each, with slope $\alpha$, on the left. 
We compare these to the slopes $1/\tslope$ determined by fitting 
$L(t)/\Lp$ as a function of $t-t_{p}$ (right, same notation). 
The resulting faint-end slope $\slope$ for each $\Lp$ and each fitting 
method are also shown; there is a clear decrease of $\slope$ with $\Lp$.
%The lower panel shows the faint-end slope $\slope$ as a function of 
%$\Lp$ for these simulations, with errors, as determined by the 
%Schechter function fits ($\times$'s) and decay of $L(t)$ (circles). 
%The resulting values of $\slope$ clearly increase with $\Lp$. 

\begin{figure*}
    \centering
    \plotone{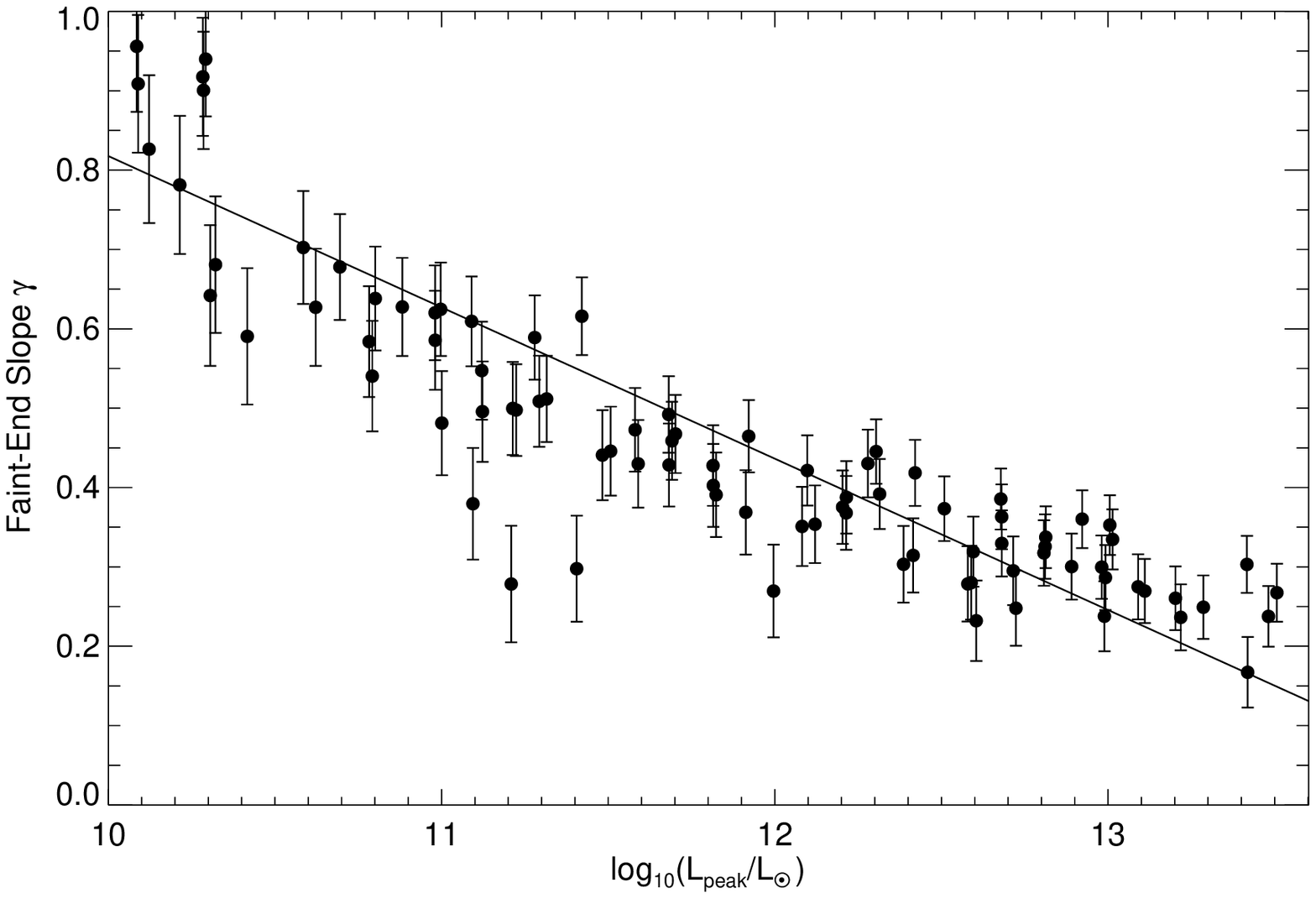}
    \caption{Faint-end QLF/quasar lifetime slope $\slope=\alpha$ determined from 
    fits to the quasar lifetime as a function of
    peak luminosity $\Lp$ in our simulations. 
    The best-fit log-linear relation (see Table~\ref{tbl:slopes}) 
    is shown (solid line).
    \label{fig:slope.v.Lp}}
\end{figure*}

\begin{deluxetable}{lcc}
\tabletypesize{\scriptsize}
\tablecaption{Faint-End Slope $\slope(\Lp)$\label{tbl:slopes}}
\tablewidth{0pt}
\tablehead{
\colhead{Determination\tablenotemark{a}} & \colhead{$\slope_{12}=\slope(\Lp=10^{12}\,L_{\sun})$} & 
\colhead{$\dslope$} 
}
\startdata
H05b, all times & $0.95\pm0.25$ & $-0.32\pm0.11$  \\
H05b, $t\gtrsim t(\Lp)$ & $0.76\pm0.20$ & $-0.30\pm0.07$  \\
H05c, all times & $0.94\pm0.20$ & $-0.33\pm0.08$  \\
H05c, $t\gtrsim t(\Lp)$ & $0.50\pm0.15$ & $-0.50\pm0.15$  \\
Schechter fitting & $0.44\pm0.02$ & $-0.21\pm0.02$ \\
$L(t)\propto t^{\tslope}$ decay & $0.69\pm0.05$ & $-0.30\pm0.04$ \\
Cumulative best-fit & $0.55\pm0.12$ & $-0.25\pm0.04$ \\
Blastwave model & $0.35\pm0.15$ & $-0.22\pm0.06$ 
\enddata
\tablenotetext{a}{H05b, H05c (Hopkins et al.\ 2005b,c) use a small 
subset of the simulations here, with H05c applying an ADAF-type correction 
at low accretion rates.}
\end{deluxetable}

We fit for $\slope(\Lp)$ in each of our suite of several hundred simulations 
with both methods described above. In all cases we find 
$\slope$ decreases with $\Lp$ in a manner similar to that 
in Figure~\ref{fig:sims.ex}. Given the scatter,  
the resulting slopes $\slope(\Lp)$ are well fitted by 
a simple log-linear relation, 
\begin{equation}
\slope(\Lp)=\slopezero + \dslope\,(\log(\Lp/L_{\sun})-12), 
\end{equation}
where $\slopezero\sim0.5$ and $\dslope\sim-0.3$. This is 
illustrated in Figure~\ref{fig:slope.v.Lp}, where the value of $\slope$ 
determined from fitting Schechter functions to the quasar lifetime 
for each of our simulations is plotted as a function of the peak 
luminosity of those simulations (circles, with errors), with the 
best-fit log-linear relation (line) 
(see also Figure~7 of Hopkins et al.\ 2005b).
The results for both fitting methods are summarized in Table~\ref{tbl:slopes}. 
We also show the results of \citet{H05b,H05c}, 
described previously. As shown in detail in \citet{H05e}, we 
find no systematic dependence on redshift, host galaxy properties, or 
any other varied parameters in the simulations. 

In our simulations, gas is expelled when feedback from 
accretion heats the surrounding gas to the point where 
it can no longer efficiently cool and rapidly unbinds it. 
Around more massive BHs (i.e.\ higher-$\Lp$ quasars), this 
``blowout'' event is progressively more violent; the higher 
luminosity of the quasar heats the gas more rapidly and to higher 
temperatures, resulting in a more violent and complete expulsion of 
gas (see also Cox et al.\ 2005, who show the gas mass fraction 
expelled increases with BH mass). This is clear 
in the representative simulations for which we fit 
$L(t-t_{p})$, which falls off much more rapidly 
in higher-$\Lp$ simulations as the density decreases and temperature 
(sound speed) increase more dramatically around the quasar. 

\subsection{A Simple Model of Quasar Feedback}
\label{sec:analytical}

To understand this trend in greater detail, we consider an 
analytical model for the quasar ``blowout'' event, i.e.\ the 
rapid heating and expulsion of gas when the quasar 
reaches a critical mass/luminosity, in agreement with the 
$M_{\rm BH}-\sigma$ relation. We can use such a model to 
predict and understand at least the late-time (post critical mass) evolution of the 
quasar accretion rate, although the early activity (which as demonstrated 
above is a significant contributor to the lifetime at low-$L$) is 
determined by the chaotic merger process and is much more difficult to 
accurately predict in simple analytical form. 
We note that this problem has been considered in greater detail in 
previous works \citep[e.g.][]{BL01,FL01,SO04}, but these 
authors have considered the large-scale structure and impact of these outflows on 
the outer regions of halos and clusters, the surrounding IGM, 
and reionization; we are instead interested in the very small 
scale, rapid expulsion of gas from the innermost BH-dominated 
accretion regions of the galaxy. Interestingly, these works suggest 
that feedback on these scales is critical in regulating the slope 
of the {\em bright} end of the QLF, which highlights the 
importance of quasar feedback on all scales in the QLF. 

In our numerical modeling \citep{DSH05} and 
most analytical models of the $M_{\rm BH}-\sigma$ relation
\citep[e.g.,][]{SR98,Fabian99,CO01,WL02}, 
feedback energy from accretion onto the 
central BH heats the surrounding gas (or momentum from 
coupling of the radiation field to dust drives a wind). In earlier 
stages of BH growth, this is a small luminosity, and the gas is able to 
efficiently cool and re-radiate this energy. However, in the peak 
merger stages, the BH growth is exponential (Eddington-limited), and 
as it rapidly grows in luminosity, a critical threshold is reached where the 
nearby gas can no longer cool efficiently in a local dynamical 
time. As a result, the gas is rapidly expelled, and what remains 
is heated to virial temperatures, preventing it from being easily accreted, and further 
accretion shuts down. The threshold for this behavior is determined by the 
local gas properties and gravitational potential of the host, but essentially 
all models of feedback driven self-regulation (owing to exceeding a 
critical energy input in of order a dynamical time) predict 
a relation between final black hole mass and virial velocity or spheroid 
velocity dispersion of roughly $M_{\rm BH}\propto V_{\rm vir}^{5}$ or 
$M_{\rm BH}\propto \sigma^{4}$, respectively. 

Because the gas in the central galaxy regions is able to efficiently cool prior to 
this self-regulation, and is then suddenly heated and driven out in a short time 
(in our simulations, this occurs over a timescale $\sim10^{7}$\,yr in massive mergers)
by an exponentially increasing luminosity, it is reasonable to model this 
driven outflow phase as a Sedov-Taylor-type blastwave \citep{Sedov46,Sedov59,Taylor50}, 
with energy injection from a point explosion 
with energy $E=\eta_{L} L_{Edd}(M_{\rm BH}^{f})\, t_{\rm dyn}$ ($\eta_{L}$ describes 
the efficiency of feedback coupling, $\approx 5\%$ in our simulations and similar in 
most analytical models of the $M_{\rm BH}-\sigma$ relation). 
A more detailed examination of this problem in e.g.\ \citet{FL01} shows that 
this is a remarkably accurate approximation to a full solution including 
radiative cooling, the pressure of the external medium, magnetic fields, and further effects.
Modeling the accretion rate with a Bondi-Hoyle \citep{bondi44,bondi52} parameterization
(as by definition we are considering times when the accretion rate falls below the 
Eddington limit), 
considered at the radius of influence of the BH, $R_{\rm BH}=G\,M_{\rm BH}\sigma^{-2}$
($\sigma$ being the spheroid velocity dispersion)
gives $L\propto M_{\rm BH}^{2}\rho(R_{\rm BH}) c_{s}^{-3}(R_{\rm BH})$, 
where $c_{s}$ is the isothermal sound speed (in our simulations, the 
effective sound speed of the multi-phase ISM is used, see Springel \& Hernquist 2003). 
As the accretion rate is rapidly falling from its 
peak during this stage, we can reasonably approximate $M_{\rm BH}=M_{\rm BH}^{f}=$constant, 
and thus $R_{\rm BH}=$constant, and we only need to calculate the time evolution of 
the gas density $\rho$ and sound speed $c_{s}$ at fixed $r=R_{\rm BH}$ to determine 
the time evolution of luminosity $L$ in this stage. We are only interested in the 
scaling of $L(t)\propto t^{-\beta}$ (where $t$ here is the time since the peak quasar luminosity or 
beginning of this ``blowout'' stage), so we could theoretically pick any fixed radius 
to evaluate these quantities, and thus the choice of the radius of influence of the BH 
as opposed to the transonic radius, for example, is not significant. 

We first consider the simplest, scale-invariant solution to this problem, in which 
we neglect the gravitational field of the BH and halo, following \citet{OM88}
in our derivation. We also ignore the consequences
of radiation as noted above, but 
this should not be a large effect in the very inner regions of the 
blastwave expansion in which we are interested (see also Murray et al.\ 2005, who 
consider in detail the coupling of BH radiation to dust and gas opacity 
and show that it produces 
a very similar $M_{BH}-\sigma$ relation and qualitative ``blowout'' behavior). 
As we need to describe $\rho(r,t)$ at small radii (relative to the galaxy 
size and final radius of the blastwave), we must consider the density gradient of the 
medium, for which we adopt the one-power approximation (OPA)
$\rho\propto r^{-k_{\rho}}$ for the ambient (pre-shocked) medium. 
There may be some non-spherical effects due to asymmetry in the 
density profile or e.g.\ bipolar outflows driven by the quasar, but the result 
is qualitatively quite similar to the spherical solution along any given 
sightline, and we furthermore expect that the chaotic interactions and short 
dynamical times in the 
central regions of the merger will rapidly isotropise the region of interest. 
For a self-similar OPA Sedov-Taylor solution, the shock radius 
$R_{s}$ expands as $R_{s}\propto t^{\eta}$ with the 
post-shock (internal) density profile $\rho = \rho_{1}\,(r/R_{s})^{l_{\rho}}$, where
$\rho_{1}\propto \rho(R_{s})$ is the density just inside the shock front. 
At fixed $r=R_{\rm BH}$, then, we have 
\begin{equation}
\rho(R_{\rm BH},t) \propto t^{-(k_{\rho}+l_{\rho})\,\eta}. 
\end{equation}
For a self-similar, energy-conserving blastwave, 
$\eta=2 / (5-k_{\rho})$, and mass conservation (coupled with the 
strong shock jump conditions;) requires 
$l_{\rho}=(6-(\geos+1) k_{\rho})/(\geos-1)$. 
Here, we denote the ratio of specific heats as $\geos$ 
to distinguish it from the faint-end QLF slope $\slope$. 
We then obtain 
\begin{equation}
(k_{\rho}+l_{\rho})\,\eta={\frac{4}{\geos-1}\Bigl(\frac{3-k_{\rho}}{5-k_{\rho}}\Bigr)}. 
\end{equation}
The pressure $P=\rho\, c_{s}^{2}$ also follows a power-law scaling (with internal 
power-law slope $l_{P}$ and external slope $k_{P}$), but with a 
more complicated result for $l_{P}$, 
\begin{equation}
l_{P}=\frac{3\geos^{2}+20\geos+1-(\geos+1)(3\geos+1)k_{\rho}}{2(\geos^{2}-1)}. 
\end{equation}
Thus the final scaling of $L(t)\propto\rho\, c_{s}^{-3}\propto \rho^{5/2} P^{-3/2}$
in this phase is given by $L\propto t^{-\beta}$ with 
\begin{equation}
\beta={\frac{10}{\geos-1}\Bigl(\frac{3-k_{\rho}}{5-k_{\rho}}\Bigr)}
-3\Bigl(\frac{k_{P}+l_{P}}{5-k_{\rho}}\Bigr).
\end{equation} 

This formalism 
has an exact, self-consistent solution corresponding to e.g.\ a blastwave 
in an isothermal sphere or wind, which should be a reasonable 
approximation to the mergers our simulations model in detail, with 
$k_{\rho}=2$ and $\geos=5/3$ (yielding $l_{\rho}=1$, $l_{P}=3$). 
This solution gives $\beta=2$, 
implying a faint-end QLF slope $\slope=1/2$. 
But the above equations 
should be a good approximation \citep{OM88} for a range of $\geos\sim1-5/3$ 
(corresponding to variability in the equation of state owing to the density and temperature 
dependence of star formation and radiative cooling) and $k_{\rho}\sim1-3$
(corresponding to e.g.\ the inner regions of a Hernquist 1990 spheroid 
profile with a central point mass). In fact, it is still exact for 
the equilibrium structure of a filled blastwave with the most natural 
$v\propto r$ velocity structure, which yields 
\begin{equation}
k_{\rho}=k_{\rho,\ \rm crit}=\frac{7-\geos}{\geos+1}, 
\end{equation}
\begin{equation}
\beta=\frac{1}{3\geos-1} \Bigl[20- \frac{9}{2}(\geos+1)\Bigr].
\end{equation}
For the reasonable range of equations of state 
$1\leq\geos\leq5/3$, this results in $2\leq\beta\leq11/2$, i.e.\ 
$0.2\lesssim\slope\lesssim0.5$, in good agreement with the range of 
$\slope$ observed and determined in our simulations. 

The scale-invariant Sedov-Taylor solution then explains well the typical values 
of $\slope$ implied by our simulations and the rapid falloff of $L(t)$ 
during the 
``blowout'' phase of quasar evolution. However, this does not immediately describe 
the dependence of $\slope$ on $\Lp$. This dependence is not contained, as might 
be presumed from the above derivation, in $\geos$, as essentially all our simulations 
have $\geos$ in the entire range $\sim1-5/3$ in their central regions and we 
specifically find no significant systematic dependence of $\slope$ or the quasar lifetime 
on the ISM gas equation of state (see also Hopkins et al.\ 2005e, Figures~4 \& 5). Rather, 
the dependence of $\slope$ on $\Lp$ is driven, naturally, by the fact that this 
problem is {\em not} precisely scale invariant or self-similar.

To illustrate the dependence on $\Lp$, consider a small change in the inital logarithmic slope 
of the density in the central regions, $k_{\rho}=k_{\rho}^{0}+\delta k_{\rho}$. 
For simplicity we expand about the $k_{\rho}^{0}=2$, $\geos=5/3$ exact 
solution, but our result is similar regardless of the choice of these parameters
(within the ranges defined above). 
This gives 
\begin{equation}
\beta = 2 - 13\delta k_{\rho}/3, 
\end{equation}
\begin{equation}
\slope = \frac{1}{2} \Bigl( 1+\frac{13}{6}\delta k_{\rho} \Bigr), 
\end{equation}
i.e.\ $d\slope/dk_{\rho}=13/12$. Thus, if the initial logarithmic 
density gradient in the inner regions is {\em steeper}, i.e.\ has a 
larger $k_{\rho}$ ($\rho_{0}\propto r^{-k_{\rho}}$), then 
$\beta$ is actually {\em shallower}. In detail, two competing effects occur. First,  
the density falls off more rapidly, allowing 
the blastwave front to propagate more rapidly
as it builds up more of its mass earlier and encounters less mass further 
from the central regions, but this 
effect is relatively weak, i.e.\ $\eta\rightarrow\eta+2\delta k_{\rho}/9$. 
The dominant effect is the alteration of the post-shock density profile. 
Because the density gradient is steeper, the propagation of the blastwave builds
up a less pronounced ``snowplow'', i.e.\ the mass buildup at the wavefront is 
less pronounced, implying a {\em flatter} post-shock density gradient by mass conservation 
(recall that $\rho$ increases with radius within the shocked region). Essentially, 
with the gas mass concentrated in the center, the shocked density profile 
evolves less dramatically, because less mass is added and the early blast is able to effectively 
redistribute more of the ultimately acquired mass. 
Given that less massive halos typically have higher concentrations, 
we can estimate the magnitude of this effect on the evolution of $\slope$ with $\Lp$. 
Estimating $dk_{\rho}/dc\sim0.1$ (where $c$ is the concentration index and we 
estimate a change $\sim10$ in $c$ changes the inner logarithmic slope at fixed 
$r$ by $\sim1$), we can determine $dc/d\log(\Lp)$ using 
$\Lp\propto M_{\rm BH}\propto M_{\rm vir}$ (e.g., Marconi \& Hunt 2003) and 
$c\approx 9 (M_{\rm vir}/10^{13}M_{\sun})^{-0.13}$ \citep{Bullock01}, 
and we find 
\begin{equation}
\dslope\sim -0.2\,\Bigl( \frac{\Lp}{10^{12}\,L_{\sun}}\Bigr)^{-0.13}, 
\end{equation}
in reasonable agreement with our measured dependence of $\slope(\Lp)$ 
in our simulations and only weakly dependent on $\Lp$. This actually predicts that 
the magnitude of 
$d\slope/d\log{\Lp}$ should be larger at low-$\Lp$ ($\sim-0.28$ at 
$\Lp\sim10^{11}\,L_{\sun}$) and smaller at high $\Lp$, 
($\sim-0.16$ at $\Lp\sim10^{13}\,L_{\sun}$)
which may occur (see Figure~\ref{fig:slope.v.Lp}), but this is most likely a
coincidence,
as our modeling of the blowout in scale invariant fashion and the application of 
a concentration parameter in such chaotic period in evolution of the merger 
are rough approximations at best. 

In this model of the quasar blowout phase, however, the approximate values for 
the rate at which accretion declines are simply determined by 
a standard Sedov-Taylor solution in a medium with a density gradient. 
The sudden injection of energy as the BH crosses a critical mass/luminosity 
threshold and the surrounding gas is no longer able to efficiently 
cool drives a strong outflow and heats the remaining gas, rapidly shutting 
down accretion. Incorporating the weak, but not negligible, effects of the 
change in concentration with mass breaks the scale invariance of this solution, 
as lower-mass systems have steeper inner density profiles, which flatten the 
evolution of the accretion rate in time and produce steeper $\slope$. Of course, 
many other effects will break the self-similarity of this problem as well -- 
a realistic gravitational potential will imply a characteristic scale length, 
and the physics of radiative cooling will likewise define fundamental physical 
scales. Even the scale-invariant solution incorporating radiative energy loss
depends on the logarithmic slope of the cooling function vs.\ density and temperature, 
but the values of these slopes depend themselves on the characteristic temperature 
of the blastwaves and change quite significantly over the mass scale of our 
simulations (heating to virial temperatures $c_{s}^{2}\sim \sigma^{2}$ implies 
temperatures $T\sim 10^{5}-10^{7}\,$K over the mass range shown in e.g.\ 
Figure~\ref{fig:slope.v.Lp}). Although we do not model the chaotic interactions 
and evolving BH mass of the early merger stages, the more 
violent torquing associated with more massive mergers can explain the more 
rapid, peaked BH evolution over a larger range in BH mass, even in early merger stages,  
generating a flatter quasar lifetime which spans a wider range in 
luminosity. 

Given these various scalings,
it is possible that our observed trend of $\slope(\Lp)$ could change or 
even reverse at very low $\Lp$, as in models of stellar winds in dwarf ellipticals 
for which lower-mass objects (lower $\mbh$) are more easily unbound
\citep[e.g.][]{MLF99}
(although this is more concerned with the large-scale binding of gas, as opposed 
to evolution in the inner accretion regions of interest in our modeling), 
but the masses/luminosities where this is likely to become important 
($M_{\rm gal}\lesssim10^{8}\,M_{\sun}$, 
i.e.\ $\Lp\lesssim10^{9}\,L_{\sun}$) are well below the break 
luminosity at any redshift, and thus will not affect our results. 
Likewise, this could occur at very large radius $r\gg a$ in any $\Lp$ 
system, but again we are not attempting to model the large-scale 
blastwave but only the evolution relevant to the small accretion regions. 

\subsection{Parameterizations of the Quasar Light Curve}
\label{sec:lightcurves}

The fits and analytical modeling above imply a useful, 
simple prescription for the quasar light curve 
as applied in semi-analytical models and other theoretical modeling 
which cannot resolve the detailed time history of individual objects as 
we can in our simulations. Generally, the quasar light curve is characterized by 
two ``modes'': a ``growing mode,'' characterized by high-Eddington ratio, rapid 
black hole growth, and a ``decaying mode,'' characterized by the nearly self-similar 
power-law falloff of the quasar luminosity as nearby gas is heated or expelled.  
The ``growing mode'' can be most simply parameterized by exponential growth at a constant 
Eddington ratio $\dot{m}$, with an exponential light curve 
$L=L(t=0)\exp{\{t/t_{Q}\}}$, where $t_{Q}=t_{S}/\dot{m}$ is the $e$-folding time 
and $t_{S}=4.2\times10^{7}\,$yr is the Salpeter time. Such black hole growth is 
expected in essentially all models of quasar activity, where plentiful fueling easily enables 
high accretion rates.

Once the quasar reaches a critical luminosity or mass (in e.g.\ accord with the 
$M-\sigma$ relation), it begins to heat and expel the surrounding gas, and the accretion rate rapidly 
falls off -- i.e.\ the light curve can be roughly parameterized as entering the ``decay mode'' described above.
Our fits to the quasar lightcurves after the ``blowout'' phase and analytical modeling of this 
phase of evolution as a driven blastwave suggest a power-law decline in the quasar lightcurve, which can 
be simple modeled as 
\begin{equation}
L(t)/L_{\rm peak}\ \approx\dot{m}\ 
= \frac{1}{1+(t/t_{Q})^{\beta}}. 
\end{equation}
$L_{\rm peak}$ here is the peak bolometric luminosity, just as the quasar enters the ``blowout'' phase, 
and $L(t)$ is the subsequent bolometric luminosity 
at time $t$ after the peak ($L=L_{\rm peak}$ at $t=0$). It is also not a bad approximation to use 
this to model both the light curve and accretion rate with 
$L(t)/L_{\rm peak}\approx\dot{m}$,  
because not much total black hole mass is accumulated in this mode. 
The equation shown assumes $\dot{m}\approx1$ at peak luminosity, as is true in most of our simulations, 
but this can easily be renormalized to any assumed constant $\dot{m}$ in the constant Eddington ratio 
``growing mode.'' The functional form of this equation is chosen such that it joins continuously with the 
constant Eddington ratio exponential light curve at $t=0$ ($L=L_{\rm peak}$, i.e.\ at the beginning of the 
``blowout'' stage), and behaves as our fitted power-laws at times large compared to the duration of the 
``blowout.'' 

The critical behavior determined from our simulation is this power-law decay of the 
quasar light curve at late times, $L(t)\propto(t/t_{Q})^{-\beta}$. We have measured $\beta$ directly in our 
simulations above, and estimated its value from simple analytical models of the quasar-driven ``blowout'' phase. 
To lowest order, a canonical value of $\beta=2$ is suggested by both our fits to the simulation light curves 
and the self-similar Sedov-Taylor solution for a quasar-driven blastwave and Bondi accretion. 
But we have also explicitly determined $\beta$ as a function of peak luminosity, which we 
inverted to determine $\gamma(L_{\rm peak})$ above. Our fits to the simulations yield 
\begin{equation}
\beta \approx \beta_{12}+\frac{d\beta}{d\log L_{\rm peak}}\,
\{\log_{10}(L_{\rm peak}/10^{12}\,L_{\odot})\}.
\label{eqn:beta.Lp}
\end{equation}
where $\beta_{12}\approx 1.7\pm0.3$ and $d\beta/d\log L_{\rm peak}\approx0.7\pm0.1$.
There is some ambiguity in $\beta$ depending on whether we are fitting 
to a given blowout or a whole quasar lifetime which includes the 
``growing'' phases (giving shallower effective $\beta$), as can be seen directly by the fact that 
different values of the faint-end quasar lifetime slope $\gamma$ are suggested by our fitting of 
the quasar lightcurves and our fitting of the quasar lifetime to a Schechter function. 
However, for most theoretical models which require a time-dependent quasar lightcurve, 
the parameterization desired is that of an individual blowout, and so these fits 
are used to determine Equation~\ref{eqn:beta.Lp} above. Furthermore, the 
``effective'' $\beta$ from fitting to the quasar lifetime is simply $1/\alpha$, which can 
be directly determined from the values of $\gamma(L_{\rm peak})$ given 
in Table~\ref{tbl:slopes}. There is a significant amount of 
scatter around this mean relation in $\beta$ (in contrast to the quite small scatter in $\alpha$) 
at a given $L_{\rm peak}$, 
$\Delta\beta\sim1$, which can easily be incorporated in theoretical models of quasar populations if desired. 
Clearly, this formula for $\beta(L_{\rm peak})$ cannot be extrapolated to arbitrary luminosities, as 
by definition $\beta>0$ always. In fact, most of the simulations which are well-fit by a very low $\beta$ 
are so because they never reach a high Eddington ratio and cause a significant ``blowout'' event, but 
rather they quiescently accrete for $\sim$Gyr at moderate ($\sim0.1$) Eddington ratios. 
In either case, this is only imporant $L_{\rm peak}\lesssim{\rm a\ few}\times10^{9}\,L_{\odot}$, which 
implies black hole masses $\lesssim10^{5}\,M_{\odot}$, well below 
the regimes where the processes we have modeled should be important (and below 
the limits to which our simulation light curves can be reliably estimated).

There is, unfortunately, significant ambiguity in the appropriate value of $t_{Q}$ to use in 
this simple parameterization of the quasar light curve. If we use Equation~\ref{eqn:beta.Lp}
to estimate the {\em cumulative} quasar lifetime at low luminosities ($L\ll L_{\rm peak}$, low enough that the 
contribution from the ``growing mode'' is negligible) and compare this with the Schechter function fits or 
direct calculation of the quasar lifetime from our simulations at these luminosities, 
there is a well-defined approximately constant 
$t_{Q}\approx t_{S}=4.2\times10^{7}\,$yr (or, if we allow $t_{Q}$ to 
vary with $\beta$ and demand that this match to the low-luminosity limit of the 
fitted Schechter function with the same $\Lp$, $t_{Q}\approx\beta t_{S}\sim10^{8}\,$yr).
However, there are several minor ``blowout'' events in our simulations, and 
usually at least two major ones (one at first passage, the other after the final 
merger). The $t_{Q}$ which we have constrained by fitting the integrated 
quasar lifetime is approximately the sum of these contributions. If we instead fit to 
$t_{Q}$ in an {\em individual} blowout, we obtain a significantly smaller 
value $t_{Q}\approx10^{7}\,$yr (as is also suggested by direct examination of Figure~\ref{fig:sims.ex}). 
The ``appropriate'' value of $t_{Q}$ is then somewhat determined by the ability of a given model to resolve separate 
starbursts triggered by e.g.\ passage of galaxies as opposed to the final merger, and 
resolution of triggering by minor mergers. 
However, our simulations constrain the range of reasonable $t_{Q}$ to $\sim10^{7}-10^{8}\,$yr, 
which is still a significantly more narrow range than the observations constrain \citep[e.g.,][]{Martini04}.
More important, to first order, $t_{Q}$
only controls the normalization of various predicted quantities and it should be easily calibrated in 
a given theoretical model by comparison to e.g.\ quasar number counts and the normalization of the luminosity function. 
Even with these uncertainties 
in the proper ``effective'' $t_{Q}$ due to the limited resolution of individual events in semi-analytical 
or cosmological models, the implied appropriate effective $t_{Q}$ for a given model is 
an interesting constraint on e.g.\ radiative efficiencies of quasars (as the $t_{Q}$ we fit scales with the $e$-folding 
time for black hole growth, and hence the radiative efficiency) and the effectiveness of quasar 
feedback. For example, a very short $t_{Q}$, especially in a model which is effectively summing over several 
different ``blowout'' events, implies very rapid expulsion or heating of gas and therefore highly efficient coupling 
of quasar energy or momentum to the surrounding gas in the ``blowout'' phase. 

The quasar lightcurve is, of course, much more complex than we have modeled here. Not only 
are there several such ``growing,'' ``blowout,'' and ``decaying'' modes in a given merger 
event, but each follows a light curve which is not trivial in detail, and our 
mixed exponential and power-law light curves are not always a good fit. Whenever possible, 
theoretical models should adopt the most accurate approximations to the quasar light curve 
available, attempting to resolve the detailed time history where it is important (for example, 
in estimating the properties and luminosity function of the faint quasar population). However, 
in many cases this is not possible or feasible, and a simple analytical approximation for the 
quasar light curve is necessary or convenient. The above simple parameterization 
describes both the growing and decaying phases of the quasar light curve with reasonable 
accuracy, and captures the critical properties of the quasar lifetime and its increase with 
decreasing luminosity and the tendency of bright peak luminosity quasars to spend 
extended periods in low-luminosity phases, and should therefore be useful in such modeling. 
This does not, however, include the effects of obscuration, which are not important (at least 
along most sightlines) in e.g.\ the late 
blowout stages, but can dramatically change the observed quasar light curve 
in various bands near peak luminosity, especially before the ``blowout'' phase \citep[e.g.,][]{H05b,H05e}.
These effects can be estimated from the parameterization of column density as a function of 
immediate and peak luminosity in \citet{H05e}, but as noted therein, these fits are only approximate
and do not accurately describe the ``blowout'' stage in which luminosity and column density both 
vary rapidly.

\section{Discussion \&\ Conclusions}
\label{sec:conclusions}

Realistic, luminosity-dependent quasar lifetimes, in which quasars spend 
more time at lower $L$, imply a novel interpretation of the 
QLF \citep{H05c}, in which the faint-end 
is composed of sources with larger $\Lp$ in early growth or sub-Eddington states. 
The faint-end slope $\slope$ is then determined by 
the slope of the quasar lifetime vs.\ $L$, for quasars with 
$\Lp\sim\lstar$, the break in the observed QLF corresponding 
directly to the {\em peak} in the distribution of quasar 
peak luminosities (BH masses) being formed at any redshift. 
We have determined $\slope$ from both a large range 
of hydrodynamical simulations (varying
BH and host galaxy properties) and from a simple analytical model 
of quasar feedback and the $\mbh$-$\sigma$ relation, and find 
that it is a monotonic decreasing function of $\Lp$ over the 
range of observed $\lstar$. 

\begin{figure*}
    \centering
    \plotone{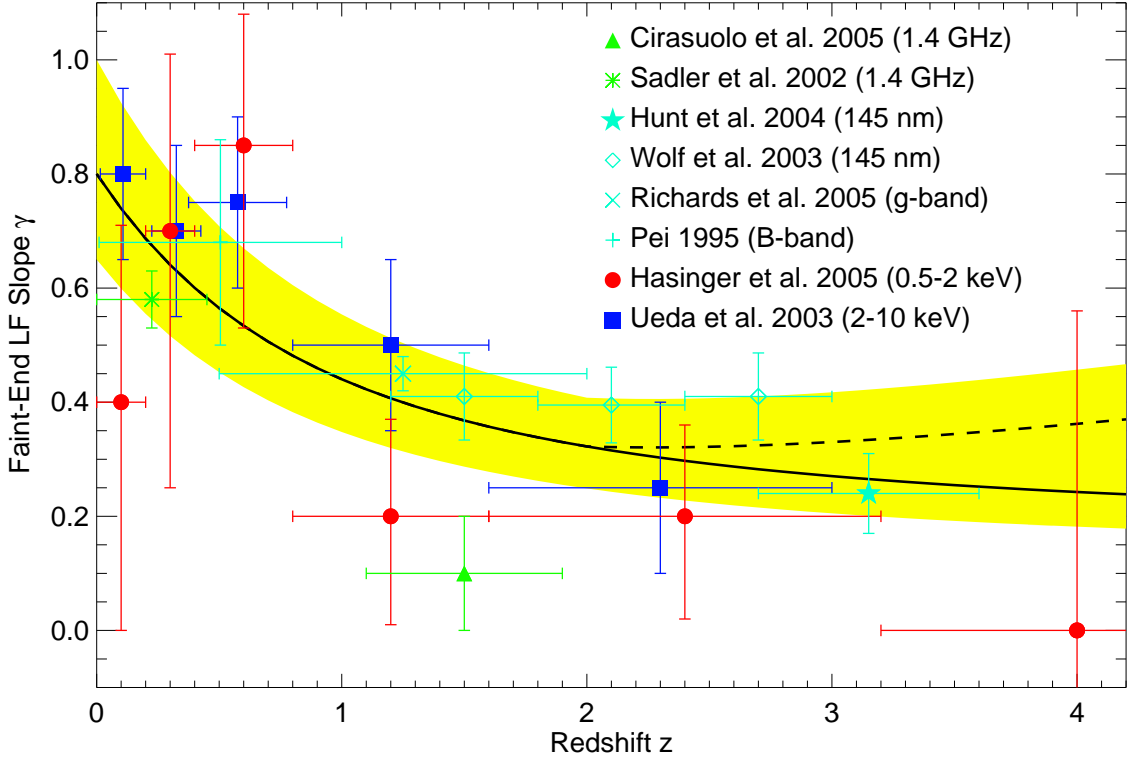}
    \caption{Predicted faint-end QLF slope $\slope$ as a function of redshift $z$, 
    from the fits in Table~\ref{tbl:slopes}. Shaded yellow shows the 
    $\sim1\sigma$ range depending on fitting method, lines the mean prediction.
    Above $z\sim2$, solid line assumes pure density 
    evolution for the QLF, dashed line assumes pure peak luminosity evolution. 
    Observations are shown (converted to bolometric QLF slopes) from different wavelengths 
    as noted. 
    \label{fig:slope.v.z}}
\end{figure*}

As the break luminosity $\lstar(z)$ is well-known from 
observations of the QLF, we use our modeling to predict $\slope(z)$. 
Figure~\ref{fig:slope.v.z} shows our result for $\slope$ in the
redshift interval $z=0-4$. Above $z\sim2-3$, the location of $\lstar$ 
becomes uncertain, and we consider both 
pure density evolution ($\lstar=$ constant or increasing, solid) 
and pure peak luminosity evolution (declining $\lstar$, dashed)
at higher redshifts. The yellow shaded range shows the 
range predicted by different fitting methods 
from Table~\ref{tbl:slopes}, lines the cumulative best-fit. 
The form of $\lstar(z)$ is taken from 
\citet{H05e,H05f}, but is based directly on fits to the observed QLFs of 
\citet{Ueda03},
\citet{Richards05}, and
\citet{HMS05}. 
We compare this to observations in the 
hard X-ray (Ueda et al.\ 2003, blue squares), soft X-ray
(Hasinger et al.\ 2005, red circles), optical 
(cyan; Wolf et al.\ 2003, diamonds; Hunt et al.\ 2004, star; 
Richards et al.\ 2005, $\times$; Pei 1995, $+$), and radio 
(green; Sadler et al.\ 2002, $\ast$; Cirasuolo et al.\ 2005, triangle). 
We convert these to bolometric luminosities and re-fit or rescale $\slope$ with 
the bolometric corrections of \citet{Marconi04}, 
which are also discussed in detail in \citet{H05d,H05e} 
and are similar to those in e.g.\ \citet{Richards05}. 
Although the uncertainties in the observed faint-end slope 
$\slope$ are large, we reproduce its value at all 
redshifts. 

\begin{figure*}
    \centering
    \plotone{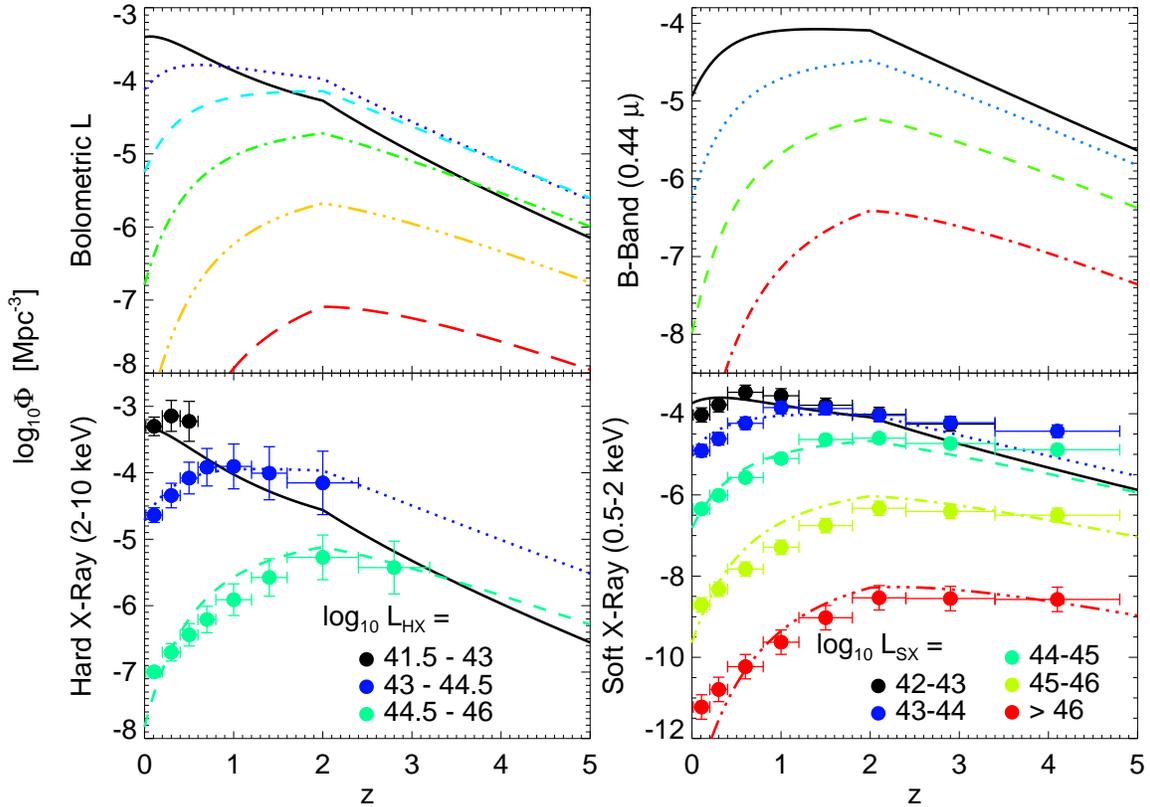}
    \caption{Predicted comoving number density in different 
    luminosity intervals as a function of redshift. 
    Upper left: Bolometric; $\log(L/L_{\sun})$ = $9-10$ (black solid line), 
    $10-11$ (blue dotted), $11-12$ (light blue short dashed), $12-13$ (green dot-dash), 
    $12-14$ (orange triple-dot-dash), $14-15$ (red long dash). 
    Upper right: B-band; $-20>M_{B}>-22.5$ (black solid), $-22.5>M_{B}>-25$ (blue dotted), 
    $-25>M_{B}>-27.5$ (green dashed), $-27.5>M_{B}>-30$ (red dot-dash). 
    Lower left: Hard X-ray; $\log(L_{HX}/{\rm erg\,s^{-1}})$ = 
    $41.5-43$ (black solid), $43-44.5$ (blue dotted), $44.5-46$ (green dashed), 
    compared to observations of \citet{Ueda03} (filled circles of corresponding color). 
    Lower right: Soft X-ray; $\log(L_{SX}/{\rm erg\,s^{-1}})$ = 
    $42-43$ (black solid), $43-44$ (blue dotted), $44-45$ (green dashed), 
    $45-46$ (yellow dot-dash), $> 46$ (red triple-dot-dash), compared to 
    observations of \citet{HMS05} (filled circles of corresponding color, normalization
    adjusted for obscuration).
    \label{fig:ldde}}
\end{figure*}

Figure~\ref{fig:ldde} demonstrates how our prediction for 
the evolution of $\slope$ with redshift translates to a ``luminosity-dependent 
density evolution'' (LDDE). 
In each panel, the integrated number density $\Phi$ (per comoving 
volume) of quasars in each of several luminosity intervals is plotted 
as a function of redshift for $z<5$. The upper left shows $\Phi$ for intervals 
in bolometric luminosity $L$: $\log(L/L_{\sun})$ = $9-10$ (black solid line), 
$10-11$ (blue dotted), $11-12$ (light blue short dashed), $12-13$ (green dot-dash), 
$12-14$ (orange triple-dot-dash), and $14-15$ (red long dash). The 
effects traditionally fitted to LDDE forms are clear; the density of higher-$L$
systems rises more rapidly to a peak at higher redshift, then falls off more slowly. 
The difference in evolution between two $L$ intervals 
becomes less dramatic with increasing $L$, as observed. 
Using the bolometric conversions of \citet{Marconi04} described above, 
we also consider the observed density evolution in other bands. 
The upper right shows density evolution in the B-band in intervals of 
B-band magnitude, $-20>M_{B}>-22.5$ (black solid), $-22.5>M_{B}>-25$ (blue dotted), 
$-25>M_{B}>-27.5$ (green dashed), $-27.5>M_{B}>-30$ (red dot-dash).
This modeling immediately demonstrates why observations in the optical 
have not found the dramatic LDDE seen in X-ray samples, 
as the effects do not become pronounced until very low luminosities not usually 
probed in B-band samples (as low as $M_{B}\gtrsim-18$), 
and the bolometric corrections 
actually slightly ``blur out'' the effect as well. 
The lower left shows our prediction for the hard X-ray, 
in three intervals of $L$(2-10 keV) = $L_{HX}$. In order to directly 
compare with the observations of \citet{Ueda03}, we adopt cgs units and 
consider the intervals $\log(L_{HX}/{\rm erg\,s^{-1}})$ = 
$41.5-43$ (black solid), $43-44.5$ (blue dotted), $44.5-46$ (green dashed). 
For each, we show the observed best-fit LDDE model with approximate
$1\sigma$ errors, in each observed redshift interval (filled circles of 
appropriate color). 
Likewise, the lower right shows our prediction for the soft X-ray 
in intervals of $L$(0.5-2 keV) = $L_{SX}$. We compare directly to 
\citet{HMS05} in the intervals $\log(L_{SX}/{\rm erg\,s^{-1}})$ = 
$42-43$ (black solid), $43-44$ (blue dotted), $44-45$ (green dashed), 
$45-46$ (yellow dot-dash), $> 46$ (red triple-dot-dash). 
Again we show the observed best-fit LDDE model in each 
observed $z$ interval (filled circles of matching color), 
but we have multiplied the QLF normalization of \citet{HMS05} by a 
factor of 10 to account for the mean obscured fraction \citep{H05e}.
For all the above, we adopt pure density evolution for $z\gtrsim2$.
We do not show the errors in our predictions, as these are actually 
dominated by the uncertainty in the fitted $\nLp$ distribution, 
not the uncertainty in $\slope(\Lp)$.

The results in Figure~\ref{fig:ldde} show that 
our modeling of the quasar lifetime predicts very accurately 
the ``luminosity-dependent density evolution'' seen in X-ray samples, 
at all luminosities and redshifts. We may slightly underpredict the 
low-luminosity, high-redshift number of sources, but this is 
unsurprising, as it is both where the observations are most 
uncertain and where, for a fixed low $L$, the break luminosity 
becomes very much larger than $L$, meaning that our modeling 
is based on the quasar lifetime at $L$ well below $\Lp$ where 
it is most uncertain. Still, the agreement is remarkable, and 
it is derived almost entirely from the quasar lifetime as a 
function of peak luminosity. The $\nLp$ distribution which 
produces the LDDE seen in Figure~\ref{fig:ldde} has a 
simple functional form: a lognormal, with constant narrow width 
and normalization, and a center/peak directly related to the 
observed break $\lstar(z)$ in the observed QLF. 
In keeping with the interpretation of the QLF from \citet{H05c}, 
in which the faint-end is made up of brighter sources 
(dominated, of course, by the peak in the $\nLp$ distribution 
at $\Lp\sim\lstar(z)$) in dimmer stages of their evolution, 
we have demonstrated that the {\em observed LDDE is entirely accounted 
for by the quasar lifetime as a function of peak luminosity, not by 
any change in the shape of the peak luminosity or black hole mass distribution}. 

A more detailed prediction of the faint-end slope 
should account for several points. Sample selection effects 
from reddening and extinction can be significant, 
especially if the fraction of obscured 
objects is a function of luminosity \citep[e.g.,][]{Ueda03,H05e}. 
Our predictions agree best with observations from the hard X-ray, 
where the effect of attenuation is minimized, but a more accurate  
prediction of $\slope$ in other wavebands must account 
for the joint evolution of obscuration and luminosity \citep{H05d,H05e}. 
To probe very low accretion rates $L\ll10^{-2}\Lp$, higher-resolution 
simulations, with more sophisticated models for low-efficiency 
accretion (rates well below Bondi or Eddington) and spectral modeling 
of the corresponding radiatively inefficient accretion flows are needed. 
Further, the contributions to the faint-end QLF from quasars with $\Lp\neq\lstar$ 
(see Hopkins et al.\ 2005e) 
will introduce overall curvature into the QLF (weakening the observed 
break, as observed by e.g.\ Wolf et al.\ 2003; Richards et al.\ 2005), which 
may aid in constraining the $\nLp$ distribution. Finally, at low enough 
$L$ (typical of LINERs and low-$L$ Seyferts), we expect
different, possibly stochastic, fueling mechanisms to contribute, which 
must be modeled to accurately predict the QLF shape below some limit. 

Still, our simple modeling of quasar feedback, and the novel 
picture of the QLF resulting from the application of realistic, luminosity-dependent 
quasar lifetimes, accurately predicts the faint-end slope of the QLF at all 
observed redshifts. Our modeling provides a simple, direct physical 
motivation for both the break luminosity $\lstar$ and faint-end slope $\slope$. 
The break $\lstar$ corresponds to the {\em peak} in the rate at which quasars 
with a given peak luminosity (final BH mass) are being created or activated
at any given time, and $\slope$ is determined (to first order) by the differential 
lifetime of these objects, as they spend substantial time at low $L\ll\Lp$, either 
in early stages of rapid BH growth to larger $L$ or in sub-Eddington states 
in transition into or out of the brief period of peak quasar activity. 
The observed $\slope$ and its evolution with redshift are a simple consequence of 
$\lstar$ and its evolution. At high-$z$, $\lstar$ is larger, implying most quasars 
have higher peak luminosity. From our simulations and analytical modeling of 
quasar feedback, we expect higher-$\Lp$ objects to both grow and expel gas more rapidly and 
violently, when they reach their final mass and feedback unbinds nearby material. 
Thus, brighter-$\Lp$ objects ``die'' more quickly, resulting in a flatter $\slope$ 
as they spend less relative time in any given $L<\Lp$. 

The observed values and evolution of $\slope$ provide a direct test of the 
model of quasar lifetimes and the QLF proposed in Hopkins et al.\ (2005a-f), 
and are not predicted by models which invoke simple ``on/off'' or pure 
exponential quasar light curves. Our model can predict 
$\slope$ as a simple function of $\lstar$, itself tied to the characteristic
quasars being created at any time, and is thus immediately useful for 
modeling the quasar contribution to reionization, which depends on 
$\slope$ at high-$z$. For purposes where $\slope$ is important, the 
quasar lifetimes fitted in \citet{H05e} should be slightly modified 
to be Schechter functions with faint-end slopes $\slope(\Lp)$ [the 
weak dependence of normalization on $\Lp$ found therein is 
entirely contained in $\slope(\Lp)$]. We have also provided simple analytical 
forms for the quasar light curve, for use in semi-analytical models and other 
theoretical models of which require the time-dependent quasar light curve (not 
simply the statistical properties contained in the quasar lifetime fits we 
have previously calculated), and which cannot resolve the detailed time 
dependence of quasar activity in individual mergers and interactions. 
We have demonstrated that the observed evolution of $\slope$ and corresponding 
``luminosity-dependent density evolution'' of the quasar population 
are not just accounted for, but actually {\em predicted} in our model 
of the QLF, as a consequence of quasar feedback being more violent in 
higher peak luminosity (larger BH mass) systems, and the fact that the 
observed faint-end QLF is dominated by sources with intrinsically brighter
peak luminosity ($\Lp\sim\lstar(z)$) in dimmer stages of their evolution. 

While more accurate
predictions require additional detailed modeling, the results presented here 
allow future observations of $\slope$ 
to directly constrain the differential quasar lifetime, even at $L\ll\Lp$, 
as our modeling shows that the faint-end QLF effectively traces the 
quasar {\em lifetime} of $\Lp\sim\lstar$ sources, {\em not} the intrinsic 
source (e.g.\ peak $L$ or BH mass) distribution. Such 
observations will further limit models 
of the distribution of quasar masses and host properties [through $\nLp$], 
and models of quasar fueling mechanisms and accretion (through the 
form of the lifetime/slope at low-$L$). As is clear in Figure~\ref{fig:slope.v.z}, the 
QLF faint-end slope as a function of redshift is still only 
poorly constrained by observations. Improved observational constraints, in particular 
samples with a uniform selection criteria which can span the range of redshifts 
$z=0-6$, for which the selection criteria could be accurately modeled and compared in 
detail with observations, 
would provide very valuable constraints on theoretical models of quasar feedback, 
e.g.\ determining the relative contributions to the faint-end QLF from 
quasars growing to larger luminosities or relaxing after their peak, and 
constraining the efficiency of coupling between accretion energy 
and the surrounding ISM. Samples which cover the relevant 
redshift range in a uniform rest frame wavelength are also especially valuable, 
and measurements in different wavelengths can provide different constraints. For example, 
the faint-end hard X-ray QLF may be more robust against obscuration effects and so provide a better indicator of 
the bolometric luminosity function, tracing e.g.\ relatively low-luminosity stages hidden 
by circumnuclear starbusts. The optical QLF, on the other hand, is in our 
modeling more closely associated with the peak quasar luminosity and ``blowout'' phase, 
meaning that the faint-end optical QLF (measurement of which would require extending the completeness of current 
high-redshift optical surveys by several magnitudes) can provide an valuable constraint on the underlying 
peak luminosity distribution, a particularly valuable quantity in itself because it determines, in our modeling, 
the QLF in all other bands, and reflects (as well as constrains) 
much more directly the underlying cosmological context, such as merger rates as a function of 
host galaxy mass. The combination of the two, yielding a reliable bolometric faint-end luminosity 
function slope and underlying peak luminosity distribution, would allow 
measurements of $\gamma(z)$ to be translated 
reliably into $\gamma(\Lp)$, and could be de-convolved to construct an 
entirely observational determination of the  
quasar lifetime as a function of luminosity. 
These observations, though difficult at very high $z$, can probe redshifts 
at which direct observations of quasar hosts and masses are inaccessible, 
constraining theories of early quasar evolution.

Combined with the modeling presented in Hopkins et al.\ (2005a-f), 
this motivates a completely self-consistent picture of the quasar 
lifetime and luminosity function, derived from the input physics of our 
simulations and without arbitrary tunable parameters. With the 
critical recognition that the quasar lifetime is {\em luminosity-dependent}, 
with quasars spending more time at low luminosities than 
their peak luminosities (as has now also been suggested by observations, 
e.g.\ Adelberger \&\ Steidel 2005), 
we have shown that a large range of observed properties of 
the quasar and spheroid population are predicted to high accuracy, over a wide 
range of observed redshifts. These works also 
provide several simple tests of our model of the quasar 
lifetime, and correspondingly the model of the faint-end slope
presented here. 
In Hopkins et al.\ (2005a-e), we explicitly demonstrate that these predictions 
include, e.g.\ the QLF at frequencies from optical to hard X-ray and at many observed redshifts, 
the distribution of Eddington ratios as a function of luminosity and redshift, 
the column density distribution at various wavelengths, the fraction of broad-line 
quasars as a function of luminosity and redshift, the active BH 
mass function of both Type I and Type II quasars, the relic BH mass function and 
total density, the cosmic X-ray background spectrum, the anti-hierarchical 
evolution of BH mass, the radio source population at high redshift, correlations between 
IR emission, star formation, and quasar obscuration, and the quasar lifetime 
as a function of luminosity and host galaxy properties. 
The distribution of Eddington ratios as a function of luminosity is a direct 
test of this model, and the degree to which typical Eddington ratios 
decrease below the break in the observed luminosity function informs the 
extent to which the faint-end slope is dominated by sources at $\Lp\sim L_{\ast}$. 
Likewise, the active BH mass function constrains the $\nLp$ distribution, and 
an accurate measurement of its peak can be used to determine the $\Lp$ which 
dominates the faint-end slope, providing a direct means to constrain the 
slope $\gamma$ as a function of $\Lp$. The broad line fraction as a function of 
luminosity can provide a similar constraint, as the luminosity at which the 
QLF transitions to being broad-line dominated is of order the turnover luminosity 
in the Schechter function fit to the quasar lifetime for $\Lp\sim$ the peak of the 
$\nLp$ distribution. 

Further, Lidz et al.\ (2005) show this model also predicts 
the correlation function and bias 
of quasars as a function of luminosity and redshift, and the quasar-galaxy 
cross-correlation function, dramatically different from the prediction in models 
with a simple quasar lifetime. The mean quasar bias (noting that the bias 
depends on luminosity much more weakly in our model than in 
simpler models of the quasar lifetime) is a direct probe of typical quasar 
host halo masses, constraining the peak in the $\nLp$ distribution and thus 
allowing for a determination of $\gamma(\Lp)$. The steepness of the 
bias vs.\ luminosity is sensitive to the width of the $\nLp$ distribution, 
and thus limits the contribution of quasars with $\Lp\neq L_{\ast}$ to 
the faint-end slope. Thus far the observations 
(e.g.\ Croom et al.\ 2005, Adelberger \& Steidel 2005) support a 
picture of a peaked $\nLp$ distribution in close agreement with our modeling, 
but the bias as a function of luminosity is still only poorly
determined by observations. 

Hopkins et al.\ (2005f) extend these predictions to the properties 
of spheroids and red galaxies, self-consistently and accurately predicting  
(from the observed {\em quasar} luminosity function) 
the distribution of velocity dispersions, the mass function, 
mass density, and star formation rates, the luminosity function in a wide range of observed 
wavebands, the total red galaxy number density and luminosity density, 
the distribution of colors as a function of magnitude and velocity dispersion
for various color-magnitude combinations, and the distribution of 
formation redshifts/ages as a function of both mass and luminosity, each
as a function of redshift at different observed redshifts. 
Our modeling also reproduces the 
X-ray emitting gas properties (Cox et al.\ 2005), morphologies 
and metallicities (Cox et al.\ 2005, in preparation), 
$M_{\rm BH}-\sigma$ relation (Di Matteo et al.\ 2005) and 
fundamental plane relation (Robertson et al.\ 2005, in preparation), 
the dispersion about each relation (Robertson et al.\ 2005), 
and bimodal color distribution (Springel et al.\ 2005a) 
of observed galaxies. The distribution of these properties and e.g.\ spheroid
luminosity and mass functions can be used to constrain the form of the 
quasar lifetime based on purely observational arguments, by using 
observed correlations between e.g.\ stellar or virial mass and BH mass 
in the context of the merger hypothesis of elliptical galaxy formation
to determine the underlying rate at which galaxies of a given 
final (combined) stellar mass must merge to produce the remnant spheroid
population and its evolution with redshift (essentially determining the 
differential growth with time and mass) and from this the 
corresponding $\nLp$ distribution. Having determined the rate at which 
quasars of a given peak luminosity are created, a de-convolution of the 
observed quasar luminosity function and $\nLp$ can constrain the 
allowed range of quasar lifetimes as a function of peak luminosity (essentially the 
reverse of our modeling, in which we de-convolve the QLF and 
quasar lifetime to determine $\nLp$).

The wide range of observations predicted by this model provides a 
number of very different means by which to observationally 
test both the qualitative picture proposed and the detailed dependence 
of quasar lifetime and corresponding faint-end QLF slope 
on quasar peak luminosity. 
We have demonstrated that a 
proper modeling of quasar evolution, 
including the critical effects of quasar feedback 
on the host galaxy in major mergers, can completely 
account for (and in fact {\em predicts}) 
the detailed evolution of the faint-end QLF slope and 
luminosity-dependent density evolution, without fitting
to the faint-end observations or fine-tuning any parameters, 
and is entirely self-consistent with the large number of detailed observations 
above, as a simple consequence of realistic, 
physically motivated, {\em luminosity-dependent} quasar lifetimes.

\acknowledgments This work 
was supported in part by NSF grants ACI 96-19019, AST 00-71019, AST
02-06299, and AST 03-07690, and NASA ATP grants NAG5-12140,
NAG5-13292, and NAG5-13381.  The simulations were performed at the
Center for Parallel Astrophysical Computing at the Harvard-Smithsonian
Center for Astrophysics.


\begin{thebibliography}{}
\bibitem[Adelberger \&\ Steidel(2005)]{Adelberger05}
Adelberger, K.~L.\ \&\ Steidel, C.~C.\ 2005, ApJ, in press [astro-ph/0505210] 
\bibitem[Barkana \& Loeb(2001)]{BL01} 
Barkana, R., \& Loeb, A.\ 2001, \physrep, 349, 125
\bibitem[Barnes \& Hernquist(1991)]{BH91}
Barnes, J.E. \& Hernquist, L. 1991, \apj, 370, L65
\bibitem[Barnes \& Hernquist(1996)]{BH96}
Barnes, J.E. \& Hernquist, L. 1996, \apj, 471, 115
\bibitem[Bondi(1952)]{bondi52}
Bondi, H. 1952, MNRAS, 112, 195
\bibitem[Bondi \& Hoyle(1944)]{bondi44}
Bondi, H. \& Hoyle, F. 1944, MNRAS, 104, 273
\bibitem[Bullock et al.(2001)]{Bullock01} 
Bullock, J.~S., Dekel, A., Kolatt, T.~S., Kravtsov, A.~V., Klypin, A.~A., Porciani, C., \& 
Primack, J.~R.\ 2001, \apj, 555, 240 
\bibitem[Ciotti \&\ Ostriker(2001)]{CO01}
Ciotti, L. \&\ Ostriker, J. P. 2001, \apj, 551, 131
\bibitem[Cirasuolo et al.(2005)]{Cirasuolo05} %faint radio LF evol. 
Cirasuolo, M., Magliocchetti, M., \& Celotti, A.\ 2005, \mnras, 357, 1267 
\bibitem[Cowie et al.(2003)]{Cowie03} 
Cowie, L.~L., Barger, A.~J., Bautz, M.~W., Brandt, 
W.~N., \& Garmire, G.~P.\ 2003, \apjl, 584, L57 
\bibitem[Cox et al.(2005)]{Cox05} % X-ray gas in remnant
Cox, T.~J., Di Matteo, T., Hernquist, L., Hopkins, P.~F., Robertson, B., \&\ 
Springel, V. 2005, \apj, submitted [astro-ph/0504156]
\bibitem[Croom et al.(2005)]{Croom05} 
Croom, S.~M., et al.\ 2005, \mnras, 356, 415 
\bibitem[Di Matteo et al.(2005)]{DSH05} % M-sigma
Di Matteo, T., Springel, V., \&\ Hernquist, L. 2005, Nature, 433, 604
\bibitem[Fabian(1999)]{Fabian99}
Fabian, A.~C. 1999, MNRAS, 308, L39
\bibitem[Fiore et al.(2003)]{Fiore03} 
Fiore, F., et al.\ 2003, \aap, 409, 79
\bibitem[Furlanetto \& Loeb(2001)]{FL01} 
Furlanetto, S.~R., \& Loeb, A.\ 2001, \apj, 556, 619
\bibitem[Hasinger et al.(2005)]{HMS05} % soft XR LF
Hasinger, G., Miyaji, T., \&\ Schmidt, M.\ 2005, \aap, in press [astro-ph/0506118]
\bibitem[Hernquist(1990)]{Hernquist90} % spheroid model
Hernquist, L. 1990, \apj, 356, 359
\bibitem[Hopkins \etal(2005a)]{H05a} % lifetimes letter
Hopkins, P.~F., Hernquist, L., Martini, P., Cox, T.~J., Robertson, B., Di Matteo, T., \&\ 
Springel, V. 2005a, \apjl, 625, L71
\bibitem[Hopkins \etal(2005b)]{H05b} % longer lifetimes/NH - optical NH distrib
Hopkins, P.~F., Hernquist, L., Cox, T.~J., Robertson, B., Di Matteo, T., Martini, P., \&\ 
Springel, V. 2005b, \apj, accepted [astro-ph/0504190]
\bibitem[Hopkins \etal(2005c)]{H05c} % QSO LF interpretation
Hopkins, P.~F., Hernquist, L., Cox, T.~J., Robertson, B., Di Matteo, T., \&\ 
Springel, V. 2005c, \apj, accepted [astro-ph/0504252]
\bibitem[Hopkins \etal(2005d)]{H05d} % NH vs. L & QSO LF (optical, softXR, hardXR)
Hopkins, P.~F., Hernquist, L., Cox, T.~J., Robertson, B., Di Matteo, T., \&\ 
Springel, V. 2005d, \apj, accepted [astro-ph/0504253]
\bibitem[Hopkins \etal(2005e)]{H05e} % Long everything-quasars
Hopkins, P.~F., Hernquist, L., Cox, T.~J., Robertson, B., Di Matteo, T., \&\ 
Springel, V. 2005e, \apj, submitted [astro-ph/0506398]
\bibitem[Hopkins \etal(2005f)]{H05f} % Galaxy properties
Hopkins, P.~F., Hernquist, L., Cox, T.~J., Robertson, B., Di Matteo, T., \&\ 
Springel, V. 2005f, \apj, submitted [astro-ph/0508167]
\bibitem[Hunt et al.(2004)]{Hunt04} % faint z=3 LF slope 
Hunt, M.~P., Steidel, C.~C., Adelberger, K.~L., \& Shapley, A.~E.\ 2004, \apj, 605, 625
\bibitem[Kauffmann \&\ Haehnelt(2000)]{KH00}
Kauffmann, G. \&\ Haehnelt, M. 2000, \mnras, 311, 576
\bibitem[La Franca et al.(2002)]{LaFranca02} 
La Franca, F., et al.\ 2002, \apj, 570, 100 
\bibitem[Lidz \etal(2005)]{Lidz05} % bias vs. L
Lidz, A., Hopkins, P.~F., Cox, T.~J., Hernquist, L., Robertson, B., Di Matteo, T., \&\ 
Springel, V. 2005, \apj, submitted [astro-ph/0507361]
\bibitem[Mac Low \& Ferrara(1999)]{MLF99} 
Mac Low, M., \& Ferrara, A.\ 1999, \apj, 513, 142 
\bibitem[Marconi \& Hunt(2003)]{MH03} 
Marconi, A., \& Hunt, L.K.\ 2003, \apj, 589, L21
\bibitem[Marconi \etal(2004)]{Marconi04} % BH mass & bol. corr.
Marconi, A., Risaliti, G., Gilli, R., Hunt, L. K., Maiolino, R., \&\ Salvati, M. 2004, \mnras, 351, 169
\bibitem[Martini(2004)]{Martini04}
Martini, P. 2004, in Carnegie Obs. Astrophys. Ser. 1, Coevolution of Black Holes
and Galaxies, ed. L.C. Ho (Cambridge: Cambridge Univ. Press), 170
\bibitem[Merloni(2004)]{Merloni04} % anti-hierarchical BH growth
Merloni, A.\ 2004, \mnras, 353, 1035
\bibitem[Mihos \& Hernquist(1996)]{MH96}
Mihos, J.C. \& Hernquist, L. 1996, \apj, 464, 641
\bibitem[Miyaji et al.(2000)]{Miyaji00} 
Miyaji, T., Hasinger, G., \& Schmidt, M.\ 2000, \aap, 353, 25
\bibitem[Miyaji et al.(2001)]{Miyaji01} 
Miyaji, T., Hasinger, G., \& Schmidt, M.\ 2001, \aap, 369, 49
\bibitem[Murray et al.(2005)]{Murray05} 
Murray, N., Quataert, E., \& Thompson, T.~A.\ 2005, \apj, 618, 569
\bibitem[Narayan \&\ Yi(1995)]{NY95} % ADAF radiative efficiency correction
Narayan, R. \&\ Yi, I. 1995, \apj, 452, 710
\bibitem[Ostriker \& McKee(1988)]{OM88} 
Ostriker, J.~P., \& McKee, C.~F.\ 1988, Reviews of Modern Physics, 60, 1 
\bibitem[Page et al.(1997)]{Page97} 
Page, M.~J., Mason, K.~O., 
McHardy, I.~M., Jones, L.~R., \& Carrera, F.~J.\ 1997, \mnras, 291, 324 
\bibitem[Pei(1995)]{Pei95} % lower-z qso LFs
Pei, Y.~C.\ 1995, \apj, 438, 623
\bibitem[Richards \etal(2005)]{Richards05} % 2SLAQ LF
Richards, G.~T.\ \etal\ 2005, in press [astro-ph/0504300]
\bibitem[Robertson \etal(2004)]{Robertson04} % Disk in cosmological sims - EOS eqn.
Robertson, B., Yoshida, N., Springel, V., \&\ Hernquist, L. 2004, \apj, 606, 32
\bibitem[Robertson \etal(2005)]{Robertson05} % M-sigma evolution
Robertson, B., Hernquist, L., Cox, T.~J., Di Matteo, T., 
Hopkins, P.~F., Martini, P., \&\ Springel, V.\ 2005, \apj, submitted [astro-ph/0506038]
\bibitem[Sadler et al.(2002)]{2002MNRAS.329..227S} % Radio z=0 LF
Sadler, E.~M., et al.\ 2002, \mnras, 329, 227
\bibitem[Scannapieco \& Oh(2004)]{SO04} 
Scannapieco, E., \& Oh, S.~P.\ 2004, \apj, 608, 62
\bibitem[Sedov(1946)]{Sedov46}
Sedov, L.~I.\ 1946, Prikl.\ Mat.\ Mekh.\ 10, 241, No.\ 2
\bibitem[Sedov(1959)]{Sedov59} 
Sedov, L.~I.\ 1959, Similarity 
and Dimensional Methods in Mechanics, New York: Academic Press, 1959
\bibitem[Silk \&\ Rees(1998)]{SR98}
Silk, J. \&\ Rees, M. J. 1998, \aap, 331, L1
\bibitem[Springel (2005)]{Springel2005}  % Gadget-2 methods
Springel, V. 2005, \mnras, submitted, [astro-ph/0505010]
\bibitem[Springel et al. (2005a)]{SDH05a} % galaxy colors
Springel, V., Di Matteo, T., \&\ Hernquist, L. 2005a, \apjl, 620, L79
\bibitem[Springel et al.(2005b)]{SDH05b} % BH + star feedback
Springel, V., Di Matteo, T., \&\ Hernquist, L. 2005b, \mnras, 361, 776
\bibitem[Springel \&\ Hernquist(2002)]{SH02} % Entropy-conserving SPH
Springel, V. \&\ Hernquist, L. 2002, \mnras, 333, 649
\bibitem[Springel \&\ Hernquist(2003)]{SH03} % Multi-phase ISM model
Springel, V. \&\ Hernquist, L. 2003, \mnras, 339, 289
\bibitem[Springel \&\ Hernquist(2005)]{SH05}
Springel, V. \&\ Hernquist, L. 2005, \apj, in press [astro-ph/0411379]
\bibitem[Taylor(1950)]{Taylor50}
Taylor, G.~I.\ 1950, Proc.\ R.\ Soc.\ London, Ser.\ A 201, 175
\bibitem[Ueda \etal(2003)]{Ueda03} % hard XR LF
Ueda, Y., Akiyama, M., Ohta, K., \&\ Miyaji, T.\ 2003, \apj, 598, 886
\bibitem[Wolf et al.(2003)]{Wolf03} % COMBO-17 LF
Wolf, C., Wisotzki, L., Borch, A., Dye, S., Kleinheinrich, M., \& Meisenheimer, K.\ 2003, \aap, 408, 499 
\bibitem[Wyithe \&\ Loeb(2002)]{WL02}
Wyithe, J. S. B. \&\ Loeb, A. 2002, \apj, 581, 886
\bibitem[Wyithe \&\ Loeb(2003)]{WL03}
Wyithe, J. S. B. \&\ Loeb, A. 2003, \apj, 595, 614

\end{thebibliography}
\end{document}